%% file: higgs2.tex
\documentclass[11pt]{article}
\usepackage{epsfig,sint,macros}
\begin{document}
\input title.tex
\input sect1.tex
\input sect2.tex
\input sect3.tex
\input concl.tex
   \bibliography{lattice}        
   \bibliographystyle{h-elsevier}   
\end{document}

%% file: title.tex
\begin{titlepage}
\begin{flushright}
  COLO-HEP-440
\end{flushright}

\vskip 1 cm
\begin{center}
  {\Large\bf Determining Lines of Constant Physics in the \\
             Confinement Phase of the SU(2) Higgs Model }
\end{center}
\vskip 1 cm
\vbox{
\centerline{
\epsfxsize=2.5 true cm
\epsfbox{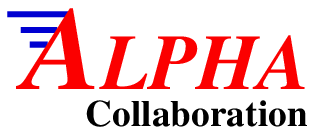}}
}
\vskip 1 cm
\begin{center}
{\large Francesco Knechtli $^{\dagger}$
}
\vskip 1.0cm
Physics Department, University of Colorado, \\
Boulder, CO 80309 USA
\end{center}
\vskip 2.5ex
{\bf Abstract}
\vskip 0.7ex
We present a method for finding lines of constant physics in the
confinement phase of the SU(2) Higgs model on the lattice.
The model is considered at finite values of the cut-off where it
behaves like an effective field theory with three independent couplings.
In particular, a renormalised quantity sensitive to a variation
of the bare Higgs quartic
self-coupling is constructed from generalised Binder cumulants.
Numerical results for the non-perturbative matching of the bare
parameters of the model between $\beta=2.2$ and $\beta=2.4$ are
presented.
  \vfill

\begin{flushleft}
  COLO-HEP-440 \\
 December 1999
\end{flushleft} 
$^{\dagger}$ {\small e-mail: knechtli@pizero.colorado.edu}
\eject

\vfill

\eject

\end{titlepage}

%% file: sect1.tex
\section{Introduction \label{s_intro}}

The conventional bare parameter space of the SU(2) 
Higgs model \cite{Montvay:1986nk} is defined by three parameters:
$\beta=4/g^2$, where $g$ is the bare gauge coupling, the hopping
parameter $\kappa$ and the bare Higgs quartic self-coupling $\lambda$. 
The continuum limit of the lattice model is at $\beta=\infty$ and
$\kappa=1/8$.
In general, the physics in the continuum limit of a lattice regularised
field theory with several relevant couplings depends on the way in which
this limit is
approached. The approach to the continuum is represented in the bare
parameter space by lines on which dimensionless and renormalised ratios $F$
of physical quantities are constant. These lines are called lines of
constant physics (LCP) \cite{Langguth:1986dr}.

It is well accepted -- supported
by the weak gauge coupling expansion
and by numerical simulations --
that the scalar part of the SU(2) Higgs model in four space-time
dimensions is a trivial theory \cite{MontMuen},
which means that in the continuum {\em limit} of the SU(2) Higgs model 
the renormalised scalar coupling $\lambda_{\rmR}$ is always zero and
therefore there are two relevant couplings only.
Nevertheless, since the coupling $\lambda$ is only marginally
irrelevant, the SU(2) Higgs model can be considered as an
{\em effective field theory} with three non-zero renormalised couplings
at finite values of the cut-off (the inverse lattice spacing
$a$). In this article, we describe a method for studying the {\em
scaling} properties of this effective field theory.
 
In practice, the LCPs in the SU(2) Higgs model
are constructed by changing $\beta$ (which corresponds to a change in the
lattice spacing) while renormalising the parameters
$\kappa$ and $\lambda$ in order to keep two independent physical
quantities $F_1$ and $F_2$ constant:
\bes\label{defLCP}
 F_i(\beta,\kappa,\lambda) & = & \mbox{constant}\,, \qquad i=1,2 \,.
\ees
Along a LCP, the value of a general physical quantity $F$
(different than $F_1$ and $F_2$) shows a dependency on the lattice
spacing like \cite{Symanzik:1983dc,Symanzik:1983gh}
\bes\label{scaling}
 F(a)|_{F_1,F_2} & = & \overline{F} +
 \rmO\left((a/\rnod)^2\right)\,,
\ees
where $\overline{F}$ is independent of $a$ and
$\rnod$ is the physical scale, like the Sommer scale
\cite{Sommer:1993ce}. The $\rmO\left((a/\rnod)^2\right)$ terms arise
from irrelevant operators and are called lattice artifacts. The absence
of $\rmO(a/\rnod)$ corrections on the right-hand side of \eq{scaling}
for a scalar theory is well accepted.
Because the renormalised coupling $\lambda_{\rmR}$ vanishes only
logarithmically along a LCP \cite{Montvay:1987dw}, i.e. like
$1/\ln(\rnod/a)$, it is possible to find a range of values of the
lattice spacing where the corrections on the right-hand side of
\eq{scaling} are negligible but $\lambda_{\rmR}$ is still sizeable. 
In this range, which we refer to as the {\em scaling} region,
for a given lattice spacing the bare coupling $\lambda$ can
be tuned such that $\lambda_{\rmR}$ assumes a particular value: the
effective field theory described by the SU(2) Higgs model with finite
cut-off has three independent couplings.

The motivation behind the work presented in this article is the
following. In a gauge theory with matter fields in the fundamental
representation of the gauge group, the potential
between a static quark and a static anti-quark\footnote{
A static quark (anti-quark) is an external source in the (complex
conjugate of the) fundamental representation of the gauge group.}
is expected to flatten at large separation of the static sources.
This phenomenon is called {\em string breaking} \cite{Schilling:Pisa99}.
Whereas string breaking has not been observed in QCD simulations yet 
\cite{Burkhalter:1998wu,Aoki:1999ff}, it has been established by numerical
simulations of the four-dimensional \cite{Knechtli:1998gf} and
three-dimensional \cite{Philipsen:1998de} SU(2) Higgs model, where a new
method for determining the static potential has been used.
The phase diagram of the SU(2) Higgs model \cite{MontMuen} is divided
into two ``phases'': the Higgs ``phase'' at large values of $\kappa$ and
the confinement ``phase''\footnote{
At small values of $\beta$ there is an analytic connection between the
two ``phases'': therefore, we put ``phase'' in quotation marks.}
at smaller $\kappa$. In the confinement
``phase'', the model has confinement properties like QCD
\cite{Montvay:1986nk,Evertz:1986vp} and is therefore suitable for
testing the method of determining the static potential.
In order to investigate the size of scaling violations in the
measurements of the static potential
a way to define LCPs in the confinement ``phase'' of the
SU(2) Higgs model has to be found. We anticipate that our results for
the static potential in the four-dimensional SU(2) Higgs model
\cite{Francesco:PhD,prep} show compatibility with scaling within tiny
errors already at moderate $\beta$ values.

In three space-time dimensions, the LCPs can be exactly
determined within perturbation theory \cite{Laine:1995ag} due to
super-renormalisability of the theory.
In the four-dimensional model, which we consider in this article,
the variations of $\kappa$ and $\lambda$ along the LCP have to be
determined {\em non-perturbatively}.
We construct two physical quantities, $F_1$ and $F_2$, which have a
sufficiently independent sensitivity on $\kappa$ and $\lambda$ in order
to be taken for the definition of a LCP.
The quantity $F_1$ is constructed from the static potential and is
sensitive to the mass of the Higgs field which determines the string
breaking distance $\rb$. The quantity $F_2$ is a generalised
Binder cumulant constructed from the connected $p$-point
functions of a zero-momentum gauge-invariant Higgs field: it
has proven to be sensitive enough to the Higgs quartic
self-coupling that we can use it to construct a LCP.
We present numerical results for the renormalisation (or matching) of the
bare parameters along a LCP between $\beta=2.2$ and $\beta=2.4$.

%% file: sect2.tex
\section{Definitions \label{s_def}}

A sensible choice for the physical quantities $F_1$ and $F_2$, which
allows us to match the parameters $\kappa$ and $\lambda$ along a LCP,
would be to
find an $F_1$ which is strongly dependent on $\kappa$ and
an $F_2$ which is strongly dependent
on $\lambda$. In this sense, a good quantity $F_1$ can be defined as
\bes\label{F1}
 F_1 & = & \rnod\,[2\mu-V_0(\rnod)]\,,
\ees
where $V_0(r)$ is the static potential and $\mu$ is the mass of a
static-light meson, the lowest bound state of a static quark and
the light dynamical Higgs field.
The difference $2\mu-V_0(r)$ is a
renormalised quantity free of self-energy contributions of the static
sources, which diverge in the continuum like $a^{-1}$.
Because the static potential flattens reaching its asymptotic value
$\lim_{r\to\infty}V_0(r)=2\mu$, the quantity $F_1$ \eq{F1} is
strictly related
to the string breaking distance $\rb$, around which the static potential
starts flattening out, and
is mainly sensitive to the value of $\mu$ and
hence to $\kappa$, which determines the mass of the Higgs field.

A remark should be added concerning finite size effects. Because physics 
shows a dependence on the physical size of the system, the latter is
part of the definition of a physical quantity $F$. Along a LCP,
the physical size of the lattice $L/\rnod$ has either to be kept
constant or at least large enough to avoid finite size effects.
As concerns the static potentials and the static-light meson spectrum,
the relevant quantities to compare are 
the lattice size $L$ and the mass gap in the gauge invariant sector of
the theory, the Higgs mass $m_{\rmH}$.
If $Lm_{\rmH}\gg1$ finite size effects are expected to
be negligible. This condition is fulfilled in our numerical computations
\cite{Francesco:PhD}.

Now, we are looking for a physical quantity $F_2$ which is sensitive to a
variation of the parameter $\lambda$, once the parameter $\kappa$ is
matched by using $F_1$ \eq{F1}. There exist perturbative equations
\cite{Montvay:1987dw} describing the change of $\lambda$ and $g$ along a 
LCP in the weak coupling limit. But the use of these equations is {\em
not} justified in our case because the bare couplings are not small
enough. The change of $\lambda$ along a LCP has to be computed in a 
non-perturbative way.
Exploratory results \cite{Montvay:1986nk} indicated that the physics of
the SU(2) Higgs model in the confinement ``phase'' is weakly dependent
on $\lambda$ once a physical quantity such as $F_1$ \eq{F1} is kept
fixed. Our studies of correlation functions from which the Higgs and
W-boson mass can be extracted confirm these observations
\cite{Francesco:PhD}.

Here, we consider a zero-momentum gauge-invariant Higgs field
\bes\label{higgszero}
 s & = &
\frac{1}{\Omega}\sum_xS(x)\;=\;\frac{1}{\Omega}\sum_x\Phid(x)\Phi(x) \,,
\ees
where $\Phi(x)$ is the complex SU(2) doublet Higgs field and $\Omega$ is the
number of lattice points. The field $S(x)$ is a composite field. The
renormalisation of composite fields is a complicated issue and we refer
to textbooks, e.g \cite{Collins,Muta}, for a detailed discussion.
The field $S(x)$ gets renormalised like
$S_{\rmR}(x)=Z_1+Z_SS(x)$. If one considers the connected $p$-point 
functions $G_{c,\rmR}(x_1,...,x_p)=\langle S_{\rmR}(x_1)\cdot\cdot\cdot
S_{\rmR}(x_p) \rangle_c$, the additive renormalisation $Z_1$ cancels and
the multiplicative renormalisation $Z_S$ is chosen such that the functions
$G_{c,\rmR}(x_1,...,x_p)$ have a well-defined continuum limit, provided
the arguments $x_1,...,x_p$ are kept at non-zero distance from one
another \cite{Luscher:1998pe}. 
\begin{figure}[tb]
\hspace{0cm}
\vspace{-1.0cm}
\centerline{\psfig{file=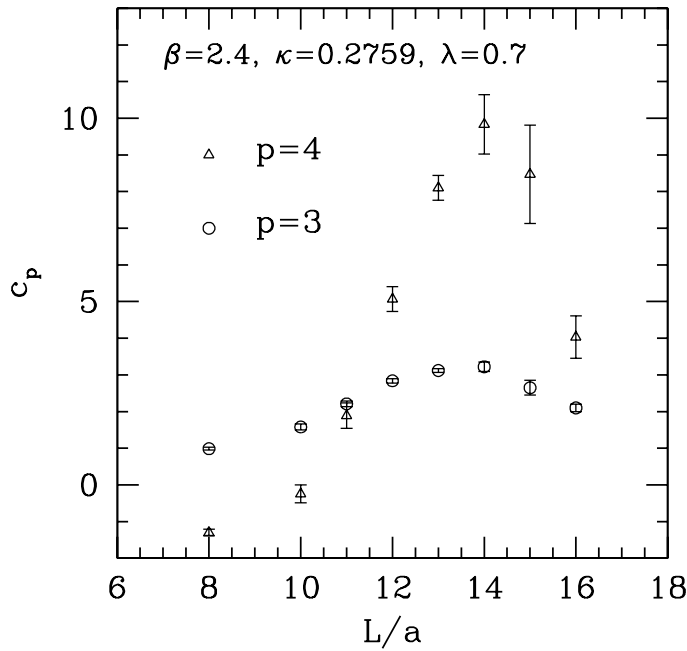,width=10cm}}
\vspace{-0.0cm}
\caption{The cumulants $c_3$ and $c_4$ defined in
\eq{cumulants} as functions of the lattice size at $\beta=2.4$.
\label{f_cumvol}}
\end{figure}

Additional care is required if some of the arguments $x_1,...,x_p$ 
in $G_{c,\rmR}(x_1,...,x_p)$ coincide, because in this case divergences 
can arise in taking the continuum limit.
This situation can be analysed in the continuum with
the help of the operator-product expansion of Wilson
\cite{Wilson:1969ey}, according to which
\bes\label{opesk}
 \langle S_{\rmR}(x_1)\cdot\cdot\cdot S_{\rmR}(x_k)\rangle_c &
 _{\mbox{$\stackrel{\displaystyle\sim}
 {\scriptstyle x_i-x_j\to\epsilon}$}} & \frac{1}{|\epsilon|^{2k}}
 \quad (\epsilon\to0)\,,
\ees
in the limit of all coincident arguments.
The power of the divergence on the right-hand side of \eq{opesk}
corresponds to the naive dimensional counting.
When we take the fields $S_{\rmR}(x)$ in \eq{opesk} at zero momentum
we get the expression
\bes\label{ppoints}
 \langle s_{\rmR}^p\rangle_c & = &
 \frac{1}{V^p} \int\rmd^4x_1\cdot\cdot\cdot\rmd^4x_p\,
 \langle S_{\rmR}(x_1)\cdot\cdot\cdot S_{\rmR}(x_p)\rangle_c
 \nonumber \\
 & = & \frac{1}{V^{p-1}}
 \int\rmd^4y_1\cdot\cdot\cdot\rmd^4y_{p-1}\,
 \langle S_{\rmR}(y_1)\cdot\cdot\cdot S_{\rmR}(y_{p-1})
 S_{\rmR}(0)\rangle_c \,,
\ees
where $V$ is the physical volume of the torus and
in the second line we used the translation invariance property
of $G_{c,\rmR}(x_1,...,x_p)$. From \eq{opesk} and naive
dimensional counting, we infer that in the limit when
$y_i\to0\;(i=1,...,p-1)$ in \eq{ppoints} the integral is finite except
in the case $p=2$ where it is naively logarithmic divergent.
This problem can be cured by defining
a modified connected 2-point function
\bes\label{2pointsreg}
 \langle\bar{s}^{(2)}_{\rmR}\rangle_c & = &
 \frac{1}{V}\int\rmd^4y\,
 \langle S_{\rmR}(y)S_{\rmR}(0)\rangle_c\,
 \sin\left(\frac{y_0}{T}\pi\right)
\ees
where $T$ is the time extension of the torus: we use $T\equiv L\equiv
V^{1/4}$.
The sine function gives a contribution proportional to $y_0$ when
$y\to0$ and therefore it makes the integral finite.
In order to be sure that {\em all} divergences associated with coincident
arguments in \eq{ppoints} are regularised for $p>2$,
it remains to consider the case of two coincident arguments,
e.g. $y_1\to y_2$. In contrast with the case $p=2$, there are still two
integrations over $y_1$ and $y_2$ and the integral is finite.
We can now define generalised Binder cumulants \cite{Binder:1981sa} $c_p$ as
\bes\label{cumulants}
 c_p & = & \frac{\langle s^p \rangle_c}{\left[\langle \bar{s}^{(2)}
   \rangle_c\right]^{p/2}} \quad (p=3,4,...) \,.
\ees
From the above discussion, it follows that the couplings $c_p$ are
well-defined and renormalised. Notice that the bare fields can be used 
in the definition of \eq{cumulants} because the multiplicative
renormalisation factors $Z_S$ cancel.

The cumulants $c_p$ can be defined on the lattice using the same
expression \eq{cumulants}: the field $s$ is given in \eq{higgszero} and
\bes\label{2ptmodified}
 \langle\bar{s}^{(2)}\rangle_c & = &
 \frac{1}{\Omega^2}\sum_{x,y}
 \langle S(x)S(y)\rangle_c\,
 \sin\left(\frac{x_0-y_0}{T}\pi\right)
\ees
is the lattice version of \eq{2pointsreg}.
In the next section, we present Monte Carlo results of the computation
of the cumulants $c_p$ on the lattice which show a linear sensitivity 
to a variation of the parameter $\lambda$, once the
physical quantity $F_1$ \eq{F1} is kept fixed.
We define the second physical quantity $F_2$ characterising the LCP to be
\bes\label{F2}
 F_2 & = & c_3|_{L/\rnod} \,,
\ees
where the physical size $L/\rnod$ of the lattice on which $c_3$ is
computed has to be kept constant along the LCP.
This is required because the cumulants $c_p$ are finite size
couplings, as we are going to see in the next section.

Our final result is that physics in the confinement ``phase'' of the
SU(2) Higgs Model can be reproduced at different values of
the lattice spacing by changing the bare parameters $\kappa$ and
$\lambda$ such that the physical quantities $F_1$ \eq{F1} and $F_2$ 
\eq{F2} are kept constant.
In the next section, we construct two points on a LCP
at $\beta$ values 2.2 and 2.4 by computing $F_1$ and $F_2$ in Monte
Carlo simulations.

%% file: sect3.tex
\section{Numerical results \label{s_num}}
\begin{table}
 \centerline{
 \begin{tabular}{|c|c|c||c|c|c|} \hline
  $\lambda$ & $\kappa$ & $\Delta\kappa$ & 
  $\lambda$ & $\kappa$ & $\Delta\kappa$ \\ \hline\hline
  0.4 & 0.2615 & 0.0001 & 0.757 & 0.2973 & 0.0001 \\ \hline
  0.5 & 0.2737 & 0.0001 & 0.96 & 0.3108 & 0.0001 \\ \hline
  0.55 & 0.2792 & 0.0002 & 1.0 & 0.3131 & 0.0001 \\ \hline
  0.7 & 0.2929 & 0.0001 & 1.5 & 0.3351 & 0.0002 \\ \hline
 \end{tabular}}
\caption{Values of $\kappa$ for different $\lambda$'s
determined at $\beta=2.2$ by using the matching
condition \eq{matching1}. The values of $\Delta\kappa$ are the
uncertainties associated with the matching. \label{t_kappa}}
\end{table}
\begin{figure}[tb]
\hspace{0cm}
\vspace{-1.0cm}
\centerline{\psfig{file=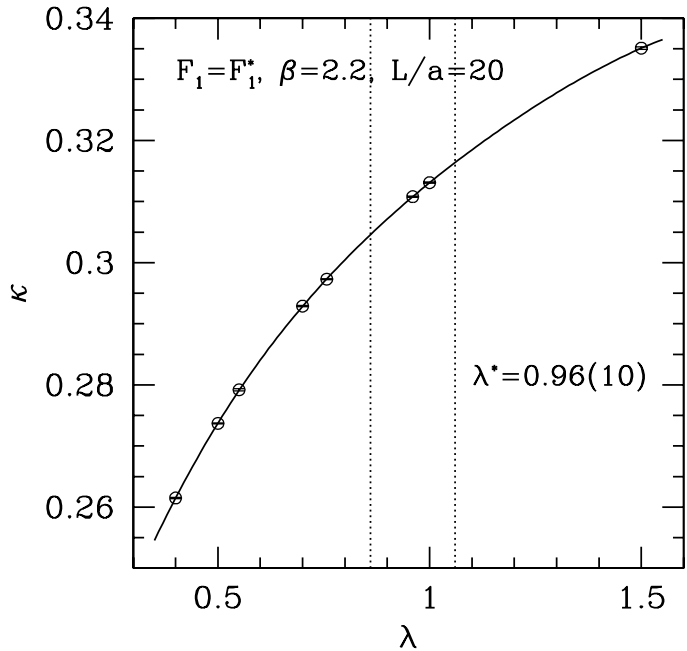,width=10cm}}
\vspace{-0.0cm}
\caption{The curve $\kappa(\lambda)$ at $\beta=2.2$ from
the matching condition \eq{matching1}. The solid line 
represents the polynomial fit \eq{fit}.
The vertical dotted band represents the matched value of $\lambda$
(with error) along the LCP \eq{LCP}. \label{f_kappa}}
\end{figure}

At $\beta=2.4$, we take the parameter set $\kappa=0.2759$, $\lambda=0.7$
\cite{Francesco:PhD} and we determine the corresponding values $F_1^*$
and $F_2^*$ (referred as matching conditions)
of the physical quantities $F_1$ \eq{F1} and $F_2$ \eq{F2}
that define the LCP.
From the quantity $F_1$, computed on a $32^4$ lattice, we obtain the
first matching condition
\bes\label{matching1}
 F_1 & = & F_1^*\,\equiv\,1.26 \,,
\ees
with a statistical error $\Delta F_1^*=0.02$. 

The cumulants $c_p$ defined in \eq{cumulants} are very small
for large values of $p$: in the Monte Carlo simulations, we were able to 
obtain a significant signal up to $p=5$. In a massive theory,
from the translation invariance property of connected $p$-point
functions (see \eq{ppoints})
the cumulants are expected to scale for large volumina $V$ of the torus like
\bes\label{cumlargev}
 c_p & \propto & \left(m_{\rmH}^4V\right)^{1-p/2} \,,
\ees
where $m_{\rmH}$ is the Higgs mass.
We computed $c_3$ and $c_4$ at $\beta=2.4$ for different lattice sizes:
the results are shown in
\fig{f_cumvol}. We observe strong variations and
therefore, a precise matching of the physical lattice size
$L/\rnod$ along the LCP is needed in order to use $F_2=c_3$ for defining 
the LCP.
We choose to compute $c_3$ on a $7^4$ lattice at $\beta=2.2$.
The physical lattice size is small enough to obtain a good signal and
the matched lattice size at $\beta=2.4$ is $L/a=13.3(3)$\footnote{
The error comes from the
determination of $\rnod/a$ at the two $\beta$ values.} in a region where
$c_3$ has no strong variations (see \fig{f_cumvol}).
Interpolating the values of $c_3$ at $\beta=2.4$ shown in 
\fig{f_cumvol} to the matched lattice size gives the second
matching condition
\bes\label{matching2}
 F_2 & = & F_2^*\,\equiv\,3.17 \,,
\ees
with an error $\Delta F_2^*=0.07$.
\begin{figure}[tb]
\hspace{0cm}
\vspace{-1.0cm}
\centerline{\psfig{file=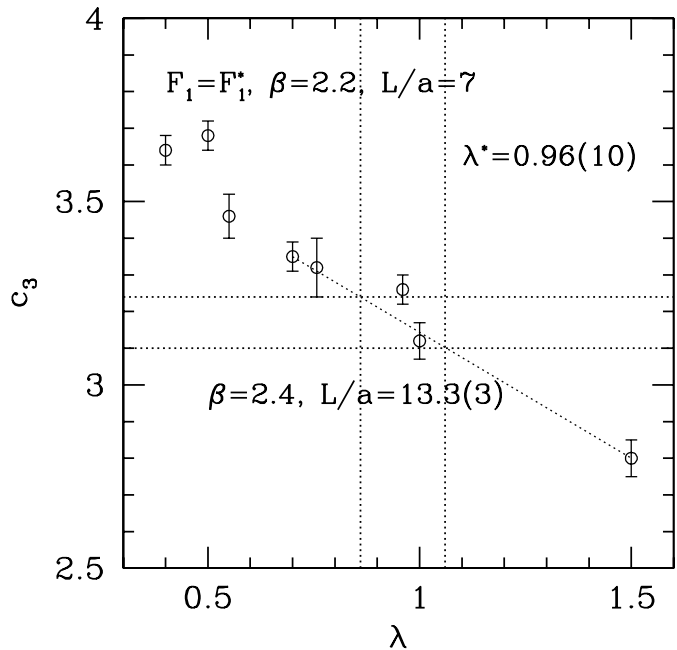,width=10cm}}
\vspace{-0.0cm}
\caption{The $\lambda$-dependence of the cumulant $c_3$ at 
$\beta=2.2$. The values for the parameter $\kappa$ are the same
as in \tab{t_kappa}.
The horizontal dotted band represents the second matching
condition \eq{matching2}. From the slope of $c_3(\lambda)$, we determine 
the matched value $\lambda^*$. \label{f_c3}}
\end{figure}

Now, we are ready to match the parameters $\kappa$ and $\lambda$ at
$\beta=2.2$. We start with matching $\kappa$ for different values of
$\lambda$ by computing $F_1$ on a $20^4$ lattice and requiring
\eq{matching1} to be fulfilled. The resulting $(\lambda,\kappa)$ pairs
are listed in \tab{t_kappa} and plotted in \fig{f_kappa}. We 
fit these points with a polynomial
\bes\label{ployfit}
 \kappa(\lambda) & = & \sum_{k=0}^M a_k(\lambda-1)^k \,.
\ees
A good description of our data is obtained for $M=4$:
\bes\label{fit}
 \kappa(\lambda) & = & 0.3131 + 0.0564\,(\lambda-1) - 0.0286\,(\lambda-1)^2
\nonumber \\ & &
+\, 0.0198\,(\lambda-1)^3 - 0.0246\,(\lambda-1)^4 \,,
\ees
and a comparison of this expression with the data points is shown in
\fig{f_kappa}. The deviation between the fitted curve \eq{fit} and the
data points is smaller than the statistical errors of the data. When
$\kappa$ is evaluated using \eq{fit}, an uncertainty of $1\cdot10^{-4}$ 
should be assigned in the range $0.4\le\lambda\le1.06$ growing up to
$4\cdot10^{-4}$ in the range $1.06<\lambda\le1.5$.
As a by-product of the simulations for the matching of $\kappa$
we obtain for the scale $\rnod$ the typical value
$\rnod/a\approx2.8$ when $F_1$ takes values near $F_1^*$. This has to be
compared with the value $\rnod/a=5.29(6)$ obtained at $\beta=2.4$: the
resulting change in the lattice spacing is
\bes\label{aratio}
 \left. \frac{a(\beta=2.2)}{a(\beta=2.4)}\right|_{F_1=F_1^*}
 & = & 1.90(4) \,,
\ees
where at $\beta=2.2$ we take $\lambda=1.0$.
\begin{figure}[tb]
\hspace{0cm}
\vspace{-1.0cm}
\centerline{\psfig{file=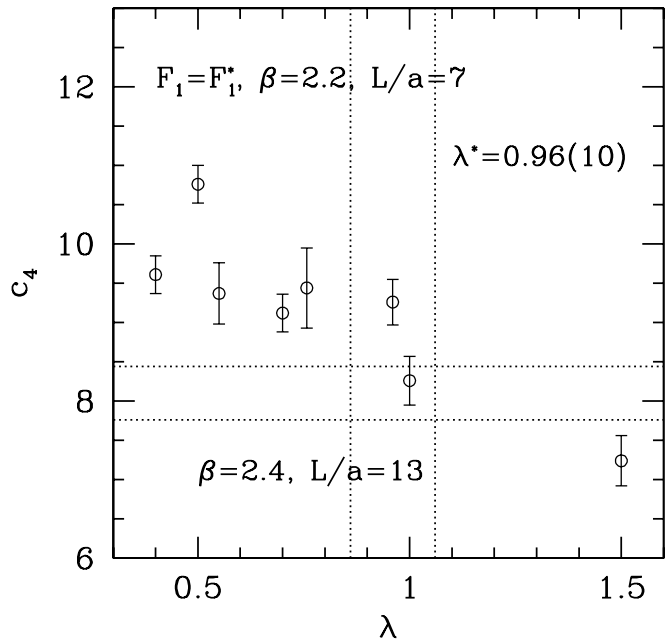,width=10cm}}
\vspace{-0.0cm}
\caption{The $\lambda$-dependence of the cumulant $c_4$ at 
$\beta=2.2$. The values for the parameter $\kappa$ are the same
as in \tab{t_kappa}.
The horizontal dotted line is the value of $c_4$ at 
$\beta=2.4$ for only approximately matched physical lattice size. The
vertical dotted line is the matched value $\lambda^*$ along the LCP
\eq{LCP}. \label{f_c4}}
\end{figure}

The matching of $\lambda$ at $\beta=2.2$ is performed by computing
$F_2=c_3$ on a $7^4$ lattice taking the parameter points
$(\lambda,\kappa)$ of \tab{t_kappa}. The results are
shown in \fig{f_c3}. The matched value
$\lambda^*$ is determined by requiring the matching condition
\eq{matching2} (represented in \fig{f_c3} by the horizontal dotted band)
to be fulfilled: we obtain $\lambda^*=0.96(10)$,
where the error is determined from the approximate slope of
$c_3(\lambda)$.
Summarising our numerical results, the points
\bes\label{LCP}
 \beta=2.2\,, & \kappa^*=\kappa(\lambda^*)\,, & \lambda^*=0.96(10) \nonumber \\
 & \mbox{and} & \\
 \beta=2.4\,, & \kappa=0.2759\,, & \lambda=0.7 \nonumber
\ees
lie on the LCP defined by the matching conditions \eq{matching1} and
\eq{matching2}. The value of $\kappa$ at $\beta=2.2$ is determined using 
the fit \eq{fit}.

In \fig{f_c4}, we show the behavior of the cumulant $c_4$ on the LCP
\eq{LCP}. The horizontal dotted band is the approximate value of $c_4$ at
$\beta=2.4$: the physical lattice size is only approximately matched.
The values of $c_4$ at $\beta=2.2$ when $\lambda$ is in the matched
interval around $\lambda^*$ are compatible with scaling.

%% file: concl.tex
\section{Conclusions \label{s_concl}}

We have described a non-perturbative method for matching the parameters
$\kappa$ and $\lambda$ along a LCP in the
confinement ``phase'' of the SU(2) Higgs model.
We consider the model at finite values of the cut-off, where
it describes an effective field theory with a
non-zero renormalised Higgs quartic self-coupling.
The lattice spacing is varied by changing $\beta$ while
two physical quantities $F_1$ and $F_2$ are kept constant. A requirement
for $F_1$ and $F_2$ is that they have a sufficiently independent 
sensitivity to $\kappa$ and $\lambda$.
The first quantity $F_1$ is derived from the
static potential and is mainly sensitive to $\kappa$. More problematic
is the definition of the second quantity $F_2$ because the physics in the
confinement ``phase'' is weakly dependent on $\lambda$.
We construct couplings, which are generalised Binder cumulants,
from the connected $p$-point
functions of a zero-momentum gauge-invariant Higgs field. In particular, 
we define a modified 2-point function in order that no divergencies
arise in taking the continuum limit.
In the range of $\lambda$ that we investigated numerically, these
couplings show a linear dependence on $\lambda$ that allows for the matching.
We determine by Monte Carlo simulations a LCP at $\beta$ values 2.2
and 2.4. Based on these results,
it is possible to study the scaling properties of physical 
quantities such as the static potential \cite{prep}.

{\bf Acknowledgement.} I thank B. Bunk, R. Sommer,
A. Hasenfratz and T. DeGrand  for helpful
discussions, the Konrad-Zuse-Zentrum {f\"ur} Informationstechnik
Berlin (ZIB) for granting CPU-resources to this project and
DESY Zeuthen, where most of this work was done.